\documentclass[a4paper]{jpconf}
\usepackage{graphicx}
\begin{document}
\title{Two particle-two hole excitations in charged current quasielastic neutrino-nucleus interactions}

\author{Marco Martini}

\address{CEA/DAM/DIF, F-91297 Arpajon, France }

\ead{marco.martini@cea.fr}

\begin{abstract}
We review the theoretical status of the models including the multinucleon emission channel in the calculation of quasielastic neutrino cross sections 
at MiniBooNE kinematics.\\
\textit{Contribution to
NUFACT 11, XIIIth International Workshop on Neutrino Factories, Super beams and Beta beams, 1-6 August 2011, CERN and University of Geneva 
(Submitted to IOP conference series).}
\end{abstract}

The MiniBooNE collaboration has recently measured the neutrino charged current (CC) quasielastic (QE) cross section 
on $^{12}$C for a neutrino beam with average energy of $788$ MeV \cite{AguilarArevalo:2010zc}. 
The comparison of these results with a prediction
based on the relativistic Fermi gas model using the standard value of the axial cut-off mass $M_A=1.03$ 
GeV$/c^2$ reveals a substantial discrepancy. 
In the same model the larger value of $M_A=1.35$ GeV$/c^2$ is needed to account for data.  
For nuclear physicists accustomed to the complexity of the nuclear system, 
this anomaly is likely to reflect the many-body aspect of the problem. 
Indeed Martini \textit{et al.} have pointed out \cite{Martini:2009uj} 
that, 
depending on the detection method, certain types of inelastic events can simulate quasielastic ones. 
As the MiniBooNE experimental set-up  defines a ``quasielastic'' event as one in which 
only a muon is detected, the ejection of a single nucleon
(a genuine quasielastic event) is only one possibility, and one
must in addition consider events involving a correlated nucleon pair from
which the partner nucleon is also ejected. This leads to the excitation of
2 particle-2 hole (2p-2h) states; 3p-3h excitations are also possible.
The inclusion in the quasielastic cross section of  
events in which several nucleons are ejected, leads to an excess over the genuine quasielastic value. 
Martini \textit{et al.} have argued  that this is the likely explanation of the anomaly 
showing that their evaluation can account for the excess in the cross section without any modification of the axial mass. 
After this suggestion, a number of articles 
have discussed the problem of multinucleon emission and whether it could account for the anomaly. 
The aim of this work is to review the actual theoretical status on this subject.

The theoretical studies of 2p-2h in connection with CCQE at MiniBooNE kinematics are actually performed 
essentially by three groups. There are the works of Martini \textit{et al.} \cite{Martini:2009uj,Martini:2010ex,Martini:arxiv}, 
the ones of Amaro \textit{et al.} \cite{Amaro:2010sd,Amaro:2011qb}
and the ones of Nieves \textit{et al.} \cite{Nieves:2011pp,Nieves:2011yp}. 
Recently it appears also a paper of Bodek \textit{et al.} \cite{Bodek:2011ps} considering 
a more phenomenological approach related to electron scattering. 
Two-nucleon emission via two-body currents is not a new idea related to neutrino-nucleus interaction, it was studied 
since the eighties in connection to the electron scattering, photon and pion absorption.
It is for example well known that the absorption of pions by nuclei is a two-nucleon mechanism. 
Concerning electron scattering, 
several theoretical calculations have shown that 2p-2h are needed to reproduce the inclusive cross section. 
In particular they contribute to fill the so called ``dip'' region between the quasielastic and the delta peaks. 
They affect the isospin spin-transverse (i.e. the magnetic) response. 
An exhaustive bibliography on these studies is beyond the purposes of this work. 
In the following we just mention the sources and the references of the calculations of the three groups mentioned above. 
The np-nh contributions in the papers of Martini \textit{et al.} are obtained starting 
from the microscopic calculations of pion absorption at threshold \cite{Shimizu:1980kb}, of pion and photon absorption 
\cite{Oset:1987re} and of the transverse 
response in electron scattering \cite{Alberico:1983zg}. They give two different parametrization of this np-nh contribution 
(which leads to practically identical neutrino total ``quasielastic'' cross section), 
one more relates to \cite{Shimizu:1980kb}, the other to \cite{Alberico:1983zg}.
In the following we present the results just obtained with this second parametrization which is characterized by a more satisfactory 
energy and momentum transfer dependence. We refer to \cite{Martini:2009uj} for details. The 2p-2h contributions considered by Amaro \textit{et al.} 
are taken from the full relativistic model of \cite{De Pace:2003xu} related to the electromagnetic responses. The work of Nieves \textit{et al.} 
can be considered as a generalization of \cite{Gil:1997}, developed for the electron scattering, to the neutrino scattering. 
The contributions related to the $\Delta$-MEC excitations are taken, as in the case of Martini  \textit{et al.} from \cite{Oset:1987re}. 
Considering these three different models, it exists some differences already of level of genuine QE which can be particularly 
important when one compares the double differential cross sections. Amaro \textit{et al.} considered the relativistic 
superscaling approach based on the superscaling behavior exhibited by electron scattering data. 
The models of Martini \textit{et al.} and Nieves \textit{et al.} are more similar: they start from a local Fermi gas picture of the nucleus. 
They consider medium polarization and collective effects through the random phase approximation (RPA) including $\Delta$-hole degrees 
of freedom, $\pi$ and $\rho$ meson exchange and $g'$ Landau-Migdal parameters in the effective $p-h$ interaction. The treatment of Nieves 
\textit{et al.} is relativistic while the original one of Martini \textit{et al.} was non relativistic. 
Recently it has been improved through relativistic corrections \cite{Martini:arxiv}. 
Concerning the 2p-2h sector we remind that there exist several contributions to two-body currents, see for example 
\cite{Nieves:2011pp,Alberico:1983zg,De Pace:2003xu}. 
There are the so called pion-in-flight term, the contact term and the $\Delta$-intermediate state or $\Delta$-MEC term. 
At level of terminology, some authors refer just to the first two terms as Meson Exchange Currents contributions, like in \cite{Martini:2009uj}, 
but actually the most current convention consists into include the $\Delta$-term into MEC. 
In order to preserve the gauge invariance of the theory also the nucleon-nucleon (NN) correlation contributions must be taken into account. 
In an infinite system they give rise to divergences. The different prescriptions to regularize them lead to a model dependence of the results. 
In this connection 
Amaro \textit{et al.} consider only the MEC contributions and not the correlations 
and the correlations-MEC interference terms. Correlations and interference are present both in Martini \textit{et al.} and Nieves \textit{et al.} 
even if Martini \textit{et al.} consider only the $\Delta$-MEC. On the other hand
the treatment of Amaro \textit{et al.} is fully relativistic as well as the one of Nieves \textit{et al.} 
while the results of Martini \textit{et al.} are related to a non-relativistic reduction of the two-body currents. 
Amaro \textit{et al.} consider the contribution only 
in the vector sector while Martini \textit{et al.} and Nieves \textit{et al.} also in the axial one. 
The virtual meson exchanged between the two nucleons is just the pion for Amaro \textit{et al.} 
while Martini \textit{et al.} and Nieves \textit{et al.} includes also $g'$ Landau-Migdal parameters. 
Nieves \textit{et al.} consider also the $\rho$ meson exchange contribution. 
\begin{figure}
\begin{minipage}[c]{55mm}
\begin{center}
      \includegraphics[width=47mm]{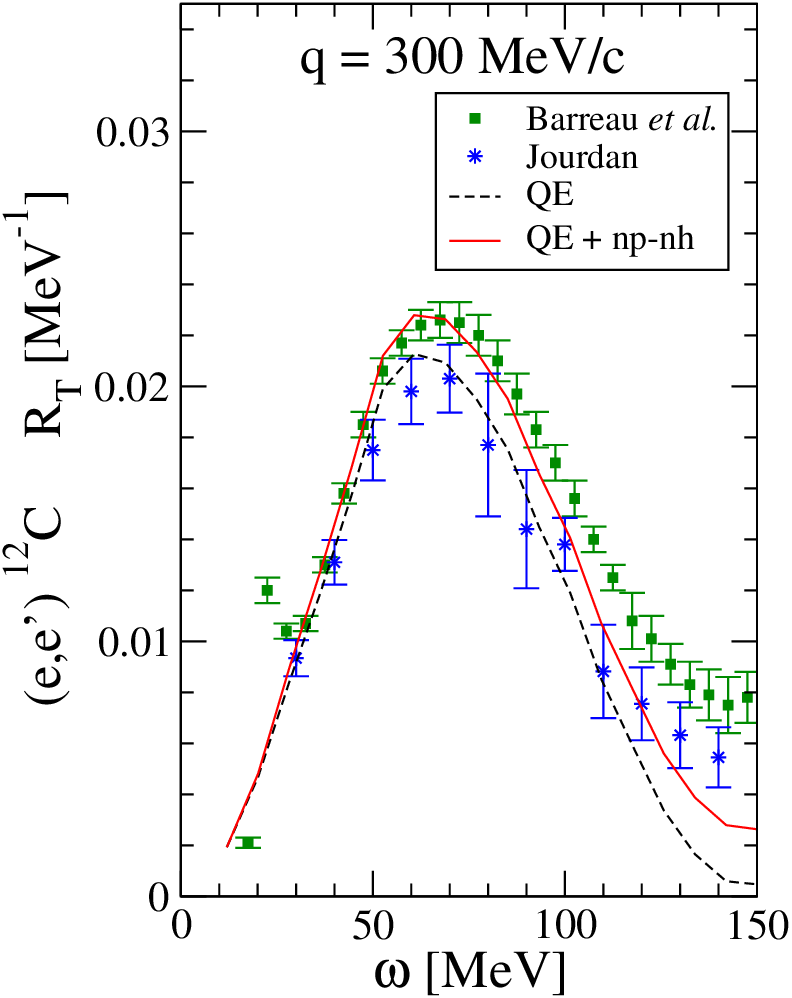} 
\end{center}
\end{minipage}
\hspace{1mm}
\begin{minipage}[c]{80mm}
      \includegraphics[width=8cm]{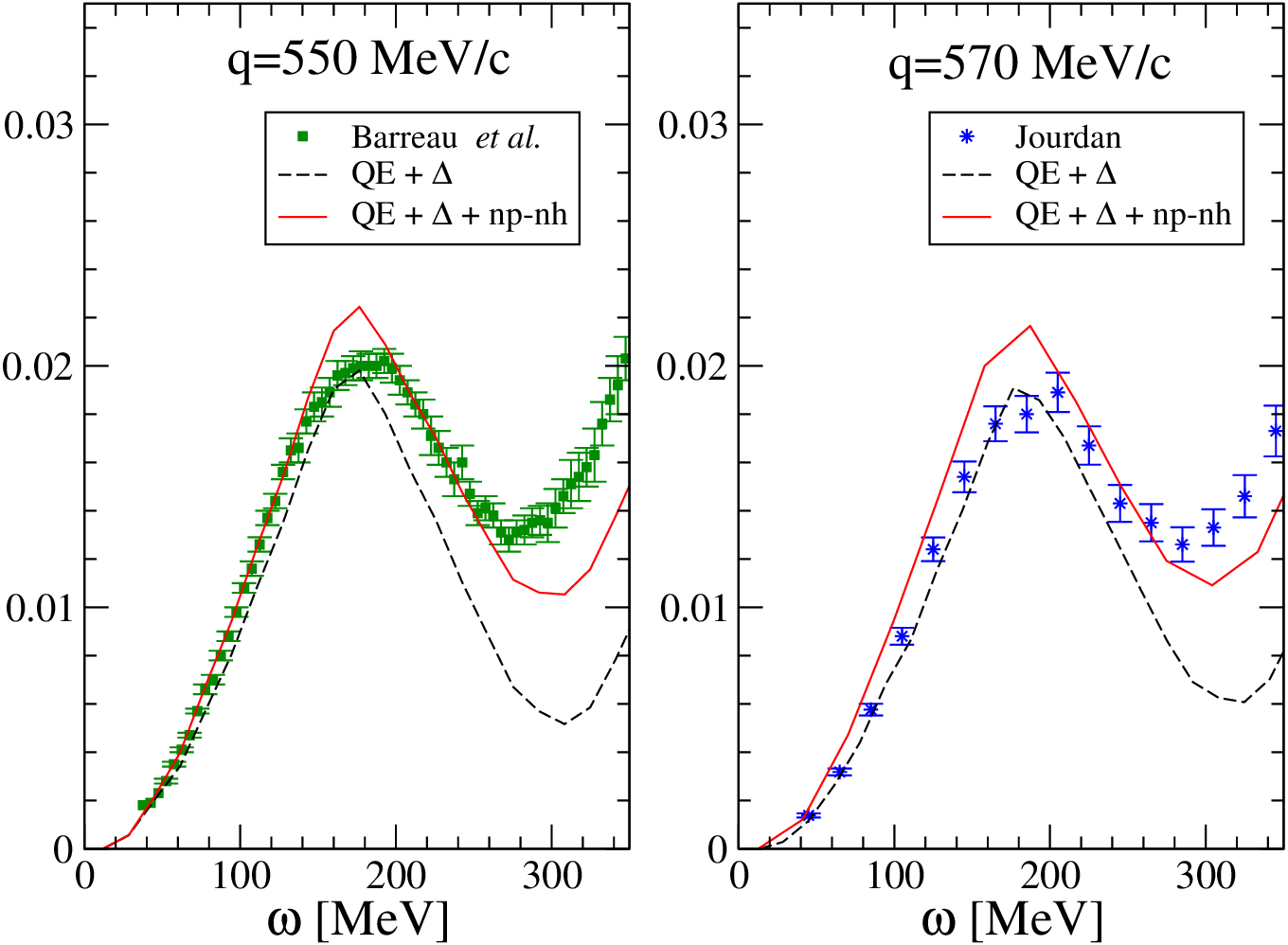}
\end{minipage}
\caption{\label{fig:rt_e} (e,e') transverse responses 
of $^{12}$C. 
Experimental data are taken from \cite{Barreau:1983ht} and \cite{Jourdan:1996ut}. 
The theoretical calculations 
are performed in the model of Martini \textit{et al.}}
\end{figure}

Before to review the results obtained in neutrino scattering, let's analyze the transverse response in electron scattering. In Fig.\ref{fig:rt_e} we present 
the transverse $(e,e')$ response on carbon calculated in the model of Martini \textit{et al.} for some values of the momentum transfer $q$ as a function of the energy 
transfer $\omega$ together with the experimental results \cite{Barreau:1983ht,Jourdan:1996ut}. 
Note that there are already some differences between the two data sets. Concerning the results at $q=300$ MeV/c 
one can compare the Martini \textit{et al.} 2p-2h contribution with the one of Fig. 41 of \cite{Gil:1997}. One can observe that they are of the same order of magnitude. At higher $q$ np-nh seems to be needed in order to reproduce data in particular in the dip region.  Bodek \textit{et al.} \cite{Bodek:2011ps} 
parametrized this enhancement 
in terms of a correction to the magnetic form factor. 
 
\begin{figure}
\begin{minipage}[c]{75mm}
      \includegraphics[width=75mm]{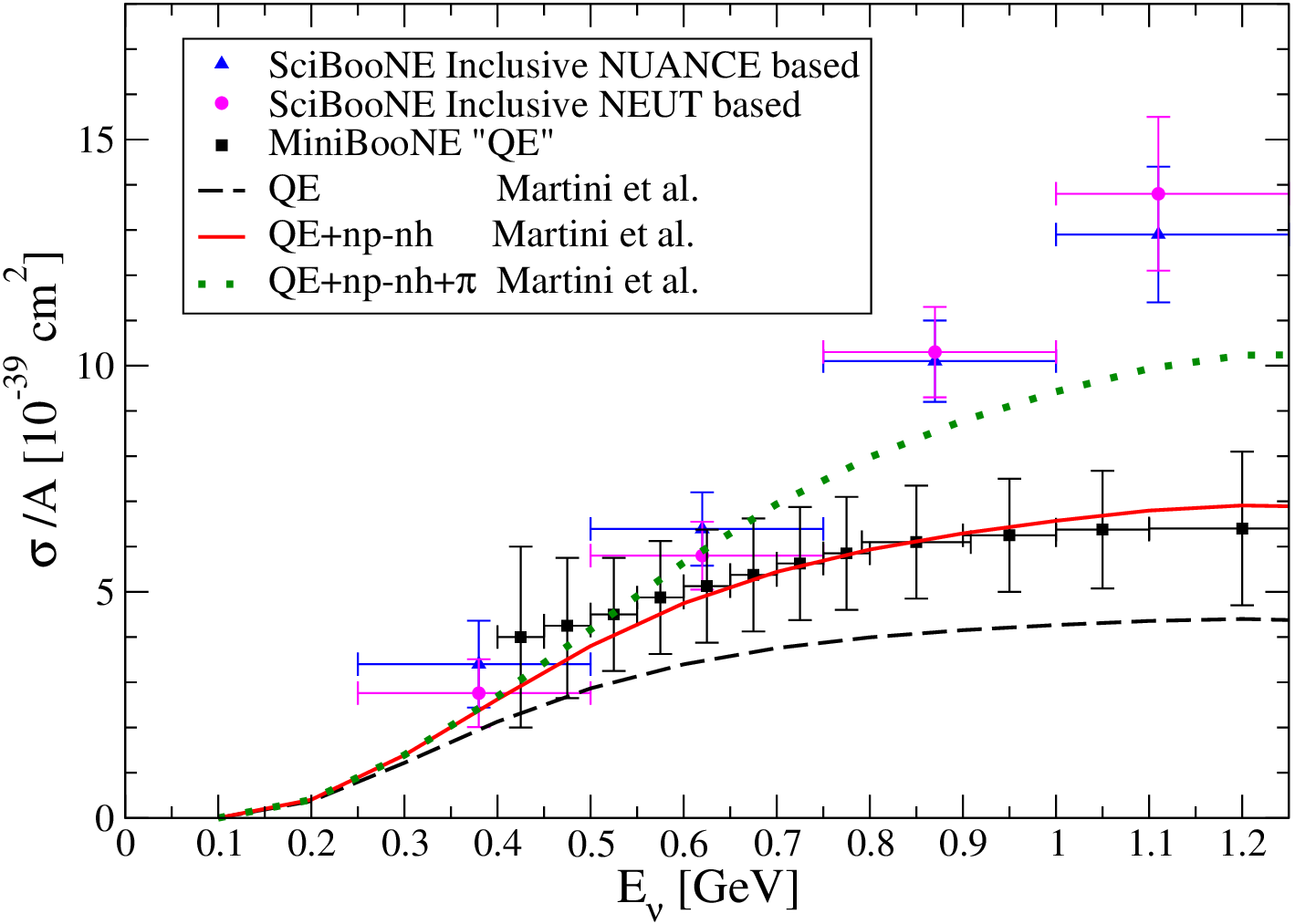} 
\caption{\label{fig_inclusive} Inclusive and ``quasielastic'' cross sections measured by SciBooNE \cite{Nakajima:2010fp} 
and MiniBooNE \cite{AguilarArevalo:2010zc} compared to Martini \textit{et al.} calculations \cite{Martini:2009uj}.}
\end{minipage}
\hspace{1mm}
\begin{minipage}[c]{75mm}
      \includegraphics[width=75mm]{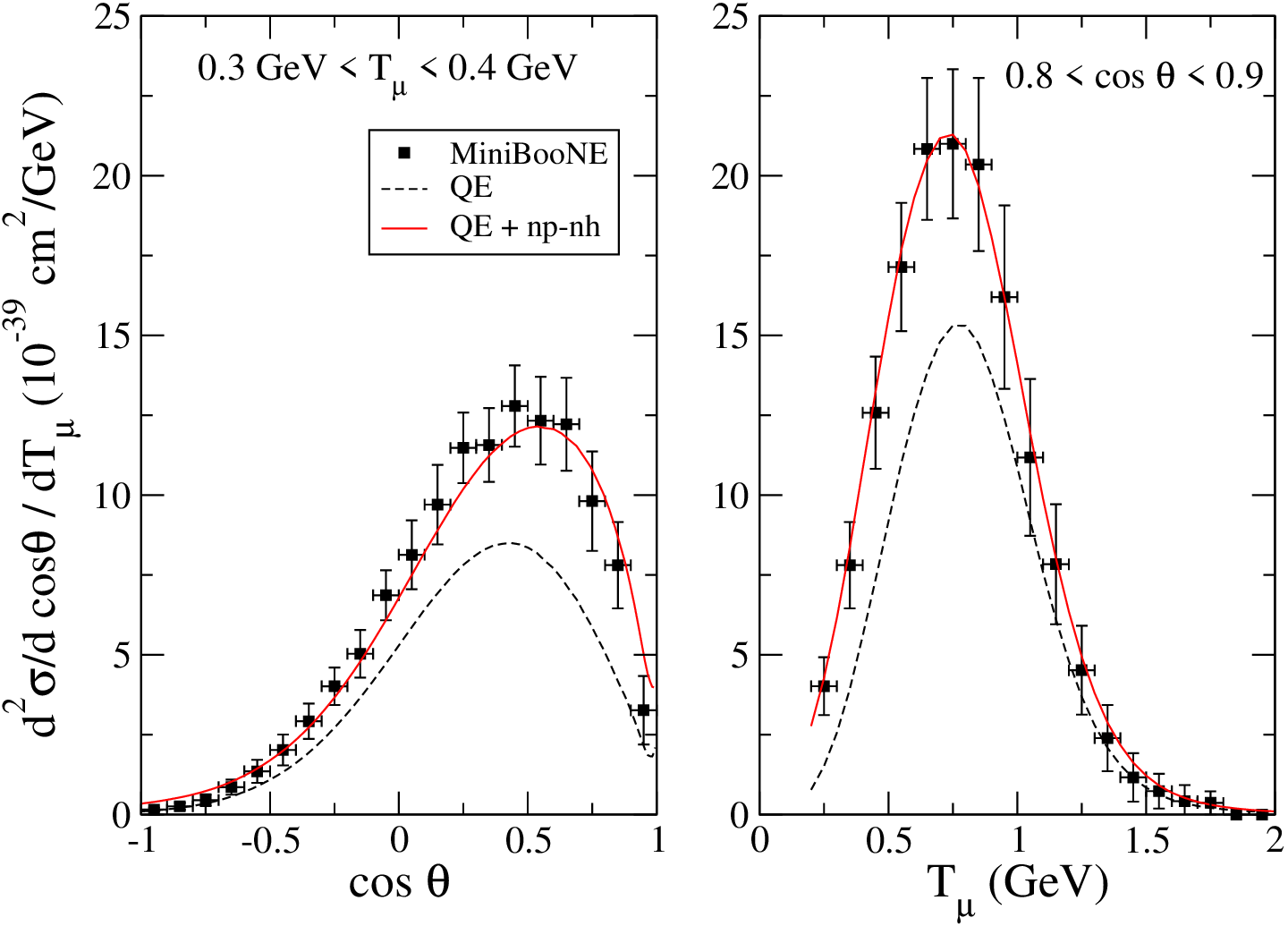}
\caption{\label{fig_d2s} MiniBooNE \cite{AguilarArevalo:2010zc} flux-averaged double differential QE cross section for some values of $T_\mu$ and cos$\theta$. 
Theoretical calculations are the ones of Martini \textit{et al.} \cite{Martini:arxiv}.}
\end{minipage}
\end{figure}

Turning to neutrino scattering, Fig.\ref{fig_inclusive} shows the ``quasielastic'' and the inclusive CC cross section as measured by MiniBooNE \cite{AguilarArevalo:2010zc} 
and SciBooNE \cite{Nakajima:2010fp} respectively.
The theoretical results of Martini \textit{et al.} are also reported. As already mentioned the inclusion of the multinucleon component 
leads to an agreement with the ``quasielastic'' MiniBooNE data. Also the inclusive cross section, which 
is less sensitive to the background subtraction with respect to exclusive channels, can be reproduced, at least up to $E_\nu \simeq 1$ 
GeV once the np-nh channel is added. Similar conclusions both for QE and inclusive cross section have been obtained 
by Nieves \textit{et al.} \cite{Nieves:2011pp}. 
Concerning the MiniBooNE ``QE'' total cross section, it can actually be reproduced without any change in the axial mass 
also by Bodek \textit{et al.} which on the other hand agrees also with NOMAD data. 
Also for Amaro \textit{et al.} the agreement with the total  ``QE'' cross section improves, at least at  high neutrino energy, 
adding to the genuine QE in the SuSA approach the MEC contribution. Nevertheless the shape of this cross section is somewhat different with respect to the other calculations, 
maybe owing to the absence of NN correlations contributions. 
The total cross section has been analyzed also in the antineutrino mode by Martini \textit{et al.}\cite{Martini:2010ex}, Nieves \textit{et al.}\cite{Nieves:2011pp} 
and  Bodek \textit{et al.}\cite{Bodek:2011ps}.
The effect of np-nh as compared to one nucleon ejection should be somewhat less important for antineutrino 
because the spin isospin response which is affected by the 2p-2h piece 
has less weight for antineutrino owing to the negative axial-vector interference. 
This effect is evident in Martini \textit{et al.} and Bodek \textit{et al.}, less in Nieves \textit{et al.} Let's turn to the MiniBooNE flux averaged ``QE'' double differential cross section. This is a directly measured quantity, hence free from the uncertainty of neutrino energy 
reconstruction which is a model dependent procedure. 
Theoretical calculations on 2p-2h contributions to this cross section have been performed by Amaro \textit{et al.}\cite{Amaro:2010sd,Amaro:2011qb}, 
Nieves \textit{et al.} \cite{Nieves:2011yp} and  
Martini \textit{et al.} \cite{Martini:arxiv}. 
Two examples of results of Martini \textit{et al.} are shown in Fig. \ref{fig_d2s}. 
For a complete comparison in all the cos$\theta$ and $T_\mu$ range measured by MiniBooNE see \cite{Martini:arxiv}. 
The agreement with data is good in all this range once the multinucleon component is incorporated. It confirm that it is possible to account 
for the MiniBooNE results without any modification of the axial mass. Similar results have already been obtained by Nieves \textit{et al.} \cite{Nieves:2011yp}. 
Although slightly lower with respect to the Martini \textit{et al.} calculations they are also fully compatible with MiniBooNE using the standard value of the axial mass.
In some kinematical regions the agreement with MiniBooNE arises in Martini \textit{et al.}, as well as in Nieves \textit{et al.}, from the opposite actions of the RPA quenching and the enhancement from the np-nh contribution. 
For Amaro \textit{et al.}\cite{Amaro:2010sd,Amaro:2011qb} the inclusion of 2p-2h MEC tends to improve the agreement with data at low angles, but it is not sufficient to 
account for discrepancy at higher angles. The absence of NN correlation contributions might be responsible for residual disagreement. 
A treatment of these contributions performed by the same group in the electron scattering sector \cite{Amaro:2010prc} seems to suggest that 
they are of the same order of magnitude as the one of 2p-2h MEC and that they are supposed to have a greater impact for large neutrino-muon angles. 
A further enhancement of the cross section expected from these contributions would lead to an agreement with MiniBooNE.

In conclusion all the actual theoretical models treating the quasielastic with the inclusion of the multinucleon emission channel 
suggest that the MiniBooNE results can be explained without any modification of the axial mass. 
The complexity of the nuclear dynamics needed to reproduce the data should be taken into account in the algorithm 
to reconstruct the neutrino energy which is crucial in neutrino oscillation experiments.
\ack
I thank Magda Ericson for useful discussions, the NUFACT11 
organizers and the WG2 conveners for the opportunity to present this review.

\end{document}